\documentclass[twocolumn,prx, superscriptaddress,longbibliography]{revtex4-1}

\usepackage{graphicx}  
\usepackage{siunitx}  
\usepackage{color}   

\begin{document}

\title{Carrier-mediated optomechanical forces in semiconductor\\ nanomembranes with coupled quantum wells}

\author{Andreas Barg}
\author{Leonardo Midolo}
\author{Gabija Kir\v{s}ansk\.{e}}
\author{Petru Tighineanu}
\author{Tommaso Pregnolato}
\affiliation{Niels Bohr Institute, University of Copenhagen, 2100 Copenhagen, Denmark}
\author{Ata\c{c} \.{I}mamo\v{g}lu}
\affiliation{Institute of Quantum Electronics, ETH Zurich, 8093 Zurich, Switzerland}
\author{Peter Lodahl}
\author{Albert Schliesser}
\author{S\o{}ren Stobbe}
\author{Eugene S. Polzik} 
\email{polzik@nbi.ku.dk}
\affiliation{Niels Bohr Institute, University of Copenhagen, 2100 Copenhagen, Denmark}

\date{\today}

\begin{abstract}

In the majority of optomechanical experiments, the interaction between light and mechanical motion is mediated by radiation pressure, which arises from momentum transfer of reflecting photons. This is an inherently weak interaction, and optically generated carriers in semiconductors have been predicted to be the mediator of different and potentially much stronger forces. Here we demonstrate optomechanical forces induced by electron-hole pairs in coupled quantum wells embedded into a free-free nanomembrane. We identify contributions from the deformation-potential and piezoelectric coupling and observe optically driven motion about three orders of magnitude larger than expected from radiation pressure. The amplitude and phase of the driven oscillations are controlled by an applied electric field, which tunes the carrier lifetime to match the mechanical period. Our work opens perspectives for not only enhancing the optomechanical interaction in a range of experiments, but also for interfacing mechanical objects with complex macroscopic quantum objects, such as excitonic condensates.

\end{abstract}

\maketitle

\section{Introduction}
Electromagnetic fields exert a force, generally known as radiation pressure, capable of displacing mechanical objects. A paradigmatic example is the reflection of a laser beam from a movable mirror arranged to form an optical cavity. This so-called cavity optomechanics approach has led to remarkable experimental demonstrations of quantum-limited sensing as well as quantum-state engineering and conversion in massive mechanical objects~\cite{AKM2014rmp,Underwood2015pra,PPKAYLR2016prl,NTMPS2016,Bagci2014n,Andrews2014np,VDWSK2012nature,LCSAT2016prl}. Since the radiation pressure force per photon is inherently weak, a lot of interest is directed towards enhancing the optomechanical interaction by improving the properties of optical and mechanical resonators, most importantly the corresponding quality factors \cite{HSK2010pra, LCLYJPV2012np,ARSAK2008photon,Norte2016prl,tsaturyan2016ultra}. Alternatively, other mechanisms than radiation pressure coupling, such as photo-thermal, electrostrictive, and piezoelectric coupling~\cite{Usami2012np, Baker2014oe,XFST2013apl, OIOSGSY2011prl, Okamoto2015nc, 2014yeonp, RG2015ultra}, have been studied using mechanical resonators made of various materials, including direct-bandgap semiconductors. These systems are promising, since they provide, even without the use of high-quality resonators, strong optomechanical interactions mediated by electronic degrees of freedom. Moreover, the systems draw on well-established methods for controlling electrical and optical properties of semiconductors, as well as the high-precision epitaxial growth of heterostructures.

\begin{figure*}[t!]
\includegraphics[width=0.65\textwidth]{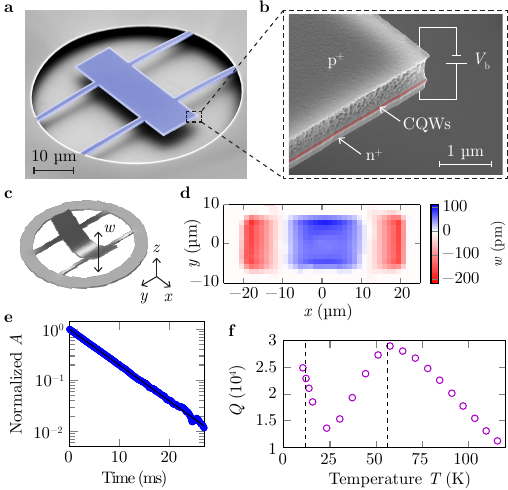}
\caption{Free-free nanomembranes with embedded CQWs. (a) False-color SEM image of the \SI{562}{\nano \meter}-thick free-free membrane (blue) and (b) zoom-in in which the vertical position of the CQWs is indicated (red). (c) Simulated mode shape and (d) spatially scanned measurement of the membrane displacement $w$ to identify the symmetric bending mode at a frequency of $\Omega_\text{m}/2\pi = \SI{1.643}{\mega \hertz}$. (e) Ringdown fitting for the symmetric bending mode at a temperature of $T = \SI{57.4}{\kelvin}$ results in a maximum quality factor of $Q = \SI{2.8e4}{}$. (f) $Q$ as a function of $T$. Statistical errors from 10 fittings are $<\SI{2.5}{\percent}$ and not shown. At $T =\SI{12}{\kelvin}$ and \SI{56}{\kelvin} (dashed lines) the linear expansion coefficient of GaAs vanishes. }
\label{fig1}
\end{figure*}

In our previous work \cite{Usami2012np} we demonstrated that the non-radiative decay of optically generated carriers in gallium arsenide (GaAs), and the subsequent heat generation, provide photo-thermal forces stronger than radiation pressure. It was also suggested that electron-hole pairs (EHPs) could induce even larger electro-mechanical interactions based on the deformation-potential~\cite{BS1950pr, CC1987prb}, which have been studied in the context of cavity optomechanics for mechanical resonators in the GHz-regime \cite{FLKJP2013prl, ABJFLLSLLSF2017prl}.
In the case of nanomembranes with lower frequencies (MHz-regime) a method to extend the carrier lifetime to match the period of oscillations needs to be devised.
In this work we present a device where such a method is implemented by embedding coupled quantum wells (CQWs) in a nanomembrane and controlling the lifetime of EHPs up to \SI{750}{\nano \s} via a bias voltage \cite{SMIPM2012prb,KTDMSLS2016prb}. Using amplitude-modulated above-bandgap laser light at a peak optical power of \SI{1}{\micro \watt}, mechanical oscillations can be driven without the use of an optical cavity to an amplitude of \SI{1.314+-0.002}{\nano \meter}, which is orders of magnitude larger than expected from radiation pressure. We demonstrate control over the amplitude and phase of the driven motion using an electric field across the CQWs. We identify two mechanisms that contribute to the carrier-induced force, namely the piezoelectric effect and the deformation-potential and distinguish them in our experiments by varying the orientation of the membranes to the crystallographic axes. 
Our findings pave the way towards optomechanical coupling based on the deformation-potential in semiconductors \cite{WRZI2004prl, Xuereb2012njp} for which the retardation mechanism necessary to realize dynamical backaction cooling and amplification \cite{MFOK2008prb}, is readily given by the delayed force.
Combining optomechanics with the rich physics of quantum dots, quantum wells, and two-dimensional electron gases~\cite{LMS2015rmp, SE2012pra, DTLS2015oe, SWHFWI2014s}, may considerably advance the development of opto-electromechanical hybrid systems.

\section{Nanomembrane with embedded coupled quantum wells}

For our experiments we fabricate free-free nanomembranes in which a rectangular plate ($\SI{40}{\micro \meter} \times \SI{12.73}{\micro \meter} \times \SI{562}{\nano \meter}$) is suspended by four thin beams \cite{Cole2011nc}. Figure \ref{fig1}a shows a scanning electron micrograph (SEM) of the device. It features top and bottom GaAs layers that are p- and n-doped, respectively, and serve as electrical contacts for the bias voltage $V_\text{b}$. InGaAs/GaAs/InGaAs 9/5/9 nm CQWs, whose band diagram is detailed in 
Ref. \cite{KTDMSLS2016prb}, are embedded between two AlGaAs barriers and placed off-center, i.e. \SI{140}{\nano \meter} above the bottom layer (Fig. \ref{fig1}b). This geometry leads to a bending mechanical force under hydrostatic stress at the CQWs, enabling an efficient coupling between carriers and a symmetric bending mode (illustrated in Fig. \ref{fig1}c) via the deformation-potential. 

The sample is placed inside a liquid-helium flow cryostat at a vacuum pressure of \SI{1e-6}{\milli \bar}. In order to detect membrane displacements we use a Michelson interferometer described in detail in Ref. \cite{Barg2017apb}. It operates at a laser wavelength of $\lambda = \SI{1064}{\nano \meter}$, which is energetically well below any interband optical transitions in the CQWs, and has therefore no influence on electronic degrees of freedom. In the probe arm of the interferometer we focus the light with an optical power of \SI{800}{\micro \watt} to a spot radius of $\sim \SI{1}{\micro \meter}$ and position the beam using a motorized $xy$-translation stage. While driving the membrane motion with a piezoelectric actuator, we measure the root mean square (RMS) amplitude $A$ and phase $\Phi$ with a lock-in amplifier at different positions on the membrane and thereby identify a symmetric bending mode at a frequency of $\Omega_\text{m}/2\pi = \SI{1.643}{\mega \hertz}$ with minimum displacement 
at the suspension points (Fig. \ref{fig1}c and \ref{fig1}d).

\begin{figure*}[tp!]
\includegraphics[width=0.65\textwidth]{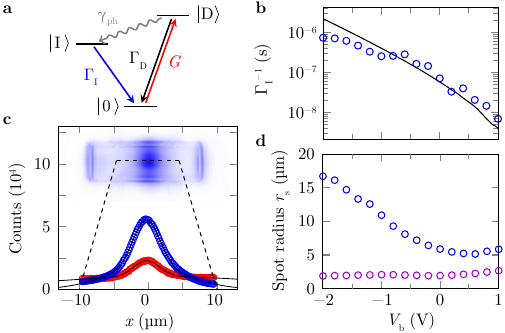}
\caption{Long-lived charge carriers in the CQWs. (a) Level scheme of the three lowest-energy EHP states and transitions. (b) Lifetime of indirect EHPs as a function of bias voltage as calculated (solid line) and extracted via exponential fits from time-resolved photoluminescence measurements (blue circles). (c) Imaged luminescence (inset) and cross-section along the horizontal dashed line for bias voltages of \SI{0}{\volt} (blue) and \SI{-2}{\volt} (red), as well as double Gaussian fits (solid lines). (d) Spot radii extracted from double-Gaussian fits versus bias voltage, where the blue and purple circles refer to the two different widths. Statistical errors from the fits in (b) and (d) are $<\SI{1}{\percent}$ and not shown.}
\label{fig2}
\end{figure*}

The measured quality factor $Q$ of the mode shows a temperature dependence with maxima around \SI{12}{\kelvin} and \SI{56}{\kelvin} (Fig. \ref{fig1}f), where the coefficient of linear expansion for GaAs and other III-V compounds used in the heterostructure is minimal~\cite{Sparks1967pr, Blakemore1982japl,adachi1992physical}. This finding reveals thermoelastic damping as a dominant damping mechanism~\cite{Okamoto2008psc,LR2000b}. We determine the quality factor via an exponential fit to the decay of the driven membrane motion (ringdown fitting) and find a maximum value of $Q = \SI{2.8e4}{}$ at a temperature of $T = \SI{57.4}{\kelvin}$ (Fig. \ref{fig1}e).

\section{Long-lived charge carrier dynamics}

The central idea behind our device is to use long-lived indirect EHPs in the CQWs as frequency-matched mediators of optomechanical forces. To model this, we describe carrier dynamics in the CQWs under illumination with above-bandgap laser light (wavelength $\lambda < \SI{890}{\nano \meter}$) as a three-level scheme \cite{DTLS2015oe} illustrated in Fig. \ref{fig2}a. We formulate rate equations for the area densities of carriers in the two lowest-energy states $n_{\text{I}}$, $n_{\text{D}}$, each dependent on time and the position $(x,y)$ on the membrane. The radiative decay rates are denote $\Gamma_{\text{I}}$ and $\Gamma_{\text{D}}$, where the indices I and D refer to indirect and direct EHP states, respectively. $\gamma_{\text{ph}}$ is a fast decay mediated by phonons. 
The EHPs are generated from (and decay into) the ground-state reservoir $\left|0\right\rangle$. 
Since the overlap of the carrier wave functions in the indirect state $\left|\text{I}\right\rangle$ is small, absorption via transition $\left|0\right\rangle \rightarrow \left|\text{I}\right\rangle$ is negligible. Therefore, carriers are only optically pumped into the direct state $\left|\text{D}\right\rangle$ with a generation rate per unit area $G = \alpha I/\hbar \omega_{\text{L}}$, where $\alpha = 0.0214$ \cite{KTDMSLS2016prb} is the absorption probability, $I$ the intensity distribution of the incident laser beam and $\omega_{\text{L}} = 2 \pi c/\lambda$. Assuming $\gamma_{\text{ph}} \gg \Gamma_{\text{D}}, \Omega$ we find the rate equations transformed into frequency domain to be $\tilde{n}_{\text{D}} \approx 0$,
\begin{equation}
	\tilde{n}_{\text{I}} \approx \frac{G}{\Gamma_{\text{I}} + i \Omega},
\label{eq:numexcit}
\end{equation}
where $\tilde{n}_{\text{D}}, \tilde{n}_{\text{I}}$ are the Fourier transforms of $n_{\text{D}}$ and $n_{\text{I}}$, respectively.
We see that all EHPs are in state $\left|\text{I}\right\rangle$ and delayed with respect to a sinusoidal optical pump $I(t) = I_0 \left( \sin\left(\Omega t\right) + 1 \right)$, where $\Omega$ is the modulation frequency and $I_0$ the amplitude. The condition $\Gamma_{\text{I}} \leq \Omega$ results in a large phase shift of $\arg{\left(\tilde{n}_\text{I} \right)} \leq -\pi/4$ which is a crucial ingredient for the dynamical backaction effects of interest here, and an increased carrier density $|\tilde{n}_\text{I}|$ resulting in a sizable mechanical force.

The lifetime $\Gamma_{\text{I}}^{-1}$ of indirect EHPs in the CQWs is tunable over two orders of magnitude by means of the bias voltage $V_\text{b}$. This feature is modeled using Fermi's golden rule, in which the carrier wave functions in the CQWs and their overlap are found by solving the one-dimensional single-particle Schr\"odinger equation numerically \cite{KTDMSLS2016prb}. Using no free parameters the model agrees well with time-resolved photoluminescence measurements (see Appendix \ref{app:plmeasurements}) at the indirect transition wavelength, as shown in Fig. \ref{fig2}b. The lifetime matches the mechanical period of $1/\Omega_\text{m} \approx \SI{100}{\nano \s}$ at $V_\text{b} \approx \SI{-0.5}{\volt}$ and increases up to a maximum value of $\Gamma_{\text{I}}^{-1} \approx \SI{750}{\nano \s}$ for $V_\text{b} = \SI{-2}{\volt}$.

\begin{figure*}[tp!]
\centering
\includegraphics[width=0.65\textwidth]{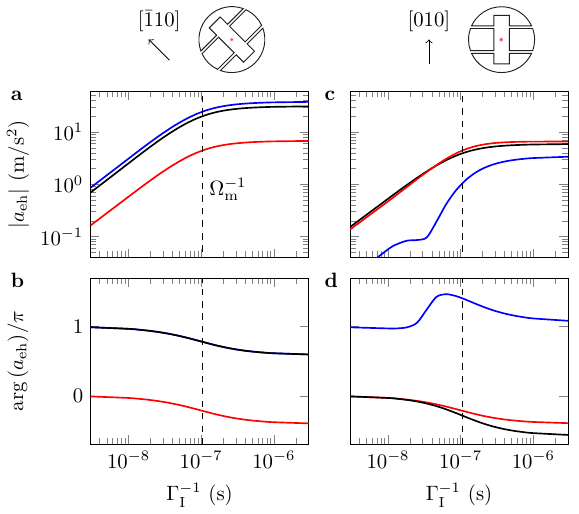}
\caption{Simulation of forces mediated by piezoelectricity and the deformation-potential. (a), (c) Amplitude and (b), (d) phase of the acceleration $a_\text{eh}(\Omega_\text{m})$, modulated at the mechanical frequency of the symmetric bending mode $\Omega_\text{m}$ versus lifetime $\Gamma_\text{I}^{-1}$ due to piezoelectricity (blue) and deformation-potential coupling (red), as well as the sum of the two (black). Dashed lines indicate where $\Gamma_\text{I}=\Omega_\text{m}$. The left column (a), (b) and the right column (c), (d) are the cases of the membrane oriented with respect to crystallographic directions $[\bar{1}10]$ and $[010]$, respectively. This is schematically shown above with topviews of the membrane pointing along the two crystal axes. The gaussian excitation spot with radius $r_\text{s} = \SI{1}{\micro\m}$ used in the simulation is shown as a red shaded area in the center.
}
\label{fig3}
\end{figure*}

With the long lifetimes achieved here, it is important to consider that EHPs in CQWs diffuse over significant distances \cite{VBSPW2005prl}. We quantify the carrier density distribution and diffusion by illuminating the center of the membrane with \SI{785}{\nano \meter} laser light, focused to a spot radius of $\sim \SI{0.75}{\micro \meter}$, and simultaneously imaging the photoluminescence from the CQWs onto a camera for different $V_\text{b}$. Directly reflected light from the sample is rejected by means of a dichroic mirror placed in front of the camera. The inset in Fig. \ref{fig2}c shows the result of a measurement at $V_\text{b} = \SI{0}{\volt}$. We find most of the photoluminescence around the excitation spot in the center of the membrane. The edges light up due to scattering of emitted photons. As shown in Fig. \ref{fig2}c, the cross-section of the image exhibits a central peak, which is best fit by a double-Gaussian function. The $1/\text{e}^2$-width (spot radius $r_\text{s}$) of one Gaussian curve is constant as a function of bias voltage and similar to the incident beam radius, while the other increases up to $r_\text{s} = \SI{16.7}{\micro \meter}$ at $V_\text{b} = \SI{-2}{\volt}$ (Fig. \ref{fig2}d). We attribute the two contributions to radiative recombination of direct and indirect EHPs, respectively.

Using the largest measured spot radii with $V_\text{b} \leq \SI{-1}{\V}$ and the corresponding lifetimes $\Gamma_\text{I}^{-1}$ we estimate the diffusion constant $D = r_\text{s}^2 \Gamma_\text{I}/2 = \SI{2.0\pm0.3}{\cm\squared\per\s}$, which is comparable to the numbers reported in similar CQW structures \cite{VBSPW2005prl}. Eq. (\ref{eq:numexcit}) can now be extended to include the diffusion of carriers as follows:
\begin{equation}
	D \nabla^2 \tilde{n}_\text{I} = (\Gamma_{\text{I}} + i \Omega) \tilde{n}_\text{I} - G.
\label{eq:diffusiondensity}
\end{equation}
We numerically solve Eq. (\ref{eq:diffusiondensity}) to find the carrier density $\tilde{n}_\text{I}$ across the membrane as a result of an excitation beam in the center with gaussian intensity distribution. The beam is modulated at the mechanical frequency $\Omega_\text{m}$ and has an amplitude of the optical power $P_0 = \pi I_0 r_\text{s}^2 /2 = \SI{0.5}{\micro\W}$. This is done for different lifetimes $\Gamma_{\text{I}}^{-1}$ and serves as a starting point for the force simulations presented below.

\section{Deformation-potential and piezoelectric forces}

There are three different forces arising from the presence of EHPs in the CQWs. First, the opposing charges in the indirect state result in a large electric field, giving rise to a shear stress described by the piezoelectric stress tensor. 
Second, the excitation of carriers into the conduction band increases the crystal lattice constant yielding a hydrostatic stress, which results in a displacement of the membrane due to the off-centered position of the CQWs. In GaAs, the hydrostatic stress is quantified by the deformation-potential $E^\text{dp} = -B(dE_{\text{g}}/dp) \approx \SI{-8}{\electronvolt}$ \cite{CC1987prb}, where $B$ is the bulk modulus, $p$ the pressure and $E_{\text{g}}$ the band gap. Third, the non-radiative relaxation of carriers induces temperature gradients in the membrane. This causes strain, known as photothermal or bolometric strain. In our experiments, this effect can be neglected since we operate at $T = \SI{12}{\kelvin}$, where the coefficient of thermal expansion for the III-V compounds used in the heterostructure is minimal.

\begin{figure*}[!]
\includegraphics[width=0.7\linewidth]{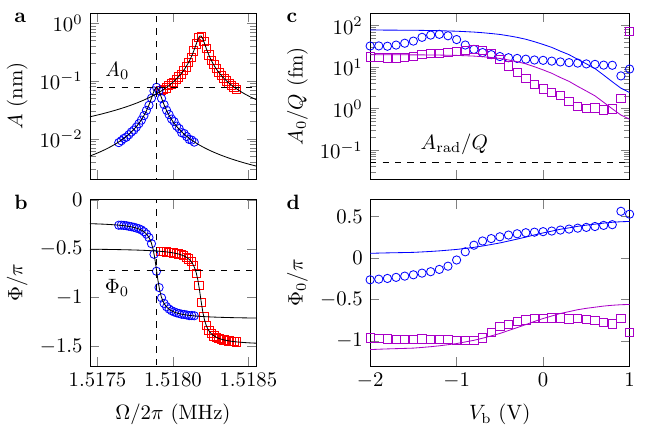}
\caption{Large optically driven motion due to tunable carrier-mediated forces. (a) RMS amplitude and (b) phase of symmetric bending mode responding to an optical drive at a wavelength of \SI{880}{\nano \meter} and incident peak power of $2 P_0 = \SI{1}{\micro \watt}$ as a function of modulation frequency. The peak RMS amplitude $A_0$ and phase $\Phi_0$ are extracted from a fit (black solid and dashed lines) for different bias voltages, here $0$ V (blue circles) and \SI{-1}{\volt} (red squares). (c) Normalized peak RMS amplitude $A_0/Q$ and (d) phase $\Phi_0$ versus bias voltage $V_\text{b}$ as measured for two different membrane orientations, $[\bar{1}10]$ (blue circles) and $[010]$ (purple squares). Errors are $<\SI{10}{\percent}$ and not shown. The solid lines are single parameter fits using our model for the total carrier-mediated force of each orientation and Eq. (\ref{eq:eqofmo}). The dashed gray line at $A_\text{rad}/Q \approx \SI{0.051}{\femto \meter}$ in (c) indicates the response expected from radiation pressure.}
\label{fig4}
\end{figure*}

Based on the calculated carrier distribution $\tilde{n}_\text{I}(x,y,\Omega_\text{m})$, we simulate the modulated acceleration $a_{\text{eh}}(\Omega_\text{m}) = F_{\text{eh}}(\Omega_\text{m})/m_{\text{eff}}$ for piezoelectric and deformation-potential coupling using finite-element analysis. Here, $F_{\text{eh}}(\Omega)$ is the corresponding force and $m_{\text{eff}}$ the effective mass of the symmetric bending mode (see Appendix \ref{app:simulations}). 
In Figure \ref{fig3}c we show the amplitude and phase of $a_{\text{eh}}(\Omega_\text{m})$ as a function of lifetime and discern two membrane orientations with respect to crystallographic axes denoted by Miller indices. If the long edges of the rectangular membrane are parallel to $[\bar110]$ the piezoelectric force dominates, while when rotated by \SI{45}{\degree} to align with $[010]$ the membrane is driven predominantly by the deformation-potential coupling. This is a direct consequence of the anisotropic character of the piezoelectric stress tensor of GaAs \cite{Sun2009spring}. Opposite signs of the total forces, revealed by the phase difference of $\pi$, for the two different membrane orientations are expected, which we can experimentally verify.
Overall, the lifetime dependence of the forces complies with the one of $\tilde{n}_\text{I}$ discussed earlier, which exhibits a phase shift and large amplitude for $\Gamma_{\text{I}} \leq \Omega_\text{m}$.

\section{Controlling large optically driven membrane motion}

To experimentally demonstrate control over the carrier-mediated forces and tunability via the bias voltage $V_\text{b}$, we perform optically driven response measurements. To this end, light from a laser diode at a wavelength of $\lambda = \SI{880}{\nano \meter}$ and an incident optical power of $2 P_0 = \SI{1}{\micro \watt}$ is amplitude-modulated by means of an acousto-optic modulator, resulting in a square wave modulation $P(t) = P_0 \left(\text{sgn}\left(\sin{\left(\Omega t\right)}\right) + 1 \right)$. Afterwards, the light is focused onto the center of a membrane, where we simultaneously detect the mechanical displacement $w(\Omega)$ using the Michelson interferometer at \SI{1064}{\nano \meter} and extract the RMS amplitude and phase with a lock-in amplifier. For different values of $V_\text{b}$, the modulation frequency $\Omega$ of the optical drive is swept in a narrow window around the mechanical eigenfrequency $\Omega_{\text{m}}$ to determine the peak RMS amplitude $A_0 = |w(\Omega_{\text{m}})|/\sqrt{2}$ and phase $\Phi_0 = \arg\left(w(\Omega_{\text{m}})\right)$ (see Fig. \ref{fig4}a and \ref{fig4}b). In addition, we perform 10 ringdown measurements and find that the quality factor $Q$ fluctuates within \SI{10}{\percent} of its mean value as a function of $V_\text{b}$. 
To compensate for this, we normalize $A_0$ with respect to $Q$. Figure \ref{fig4}c and \ref{fig4}d show the results of our measurements, confirming that the membrane response to the optical drive is significantly altered by $V_\text{b}$.

To compare the response data with our model, we assume that the displacement $w$ is mainly driven by the sum of the two modulated optomechanical forces $F_{\text{eh}}(\Omega_{\text{m}})$ considered in Fig. \ref{fig3}. With this, the following equation of motion applies:
\begin{equation}
	w(\Omega_{\text{m}}) = \chi_{\text{m}}(\Omega_{\text{m}}) \eta F_{\text{eh}}(\Omega_{\text{m}}),
\label{eq:eqofmo}
\end{equation}
where $\chi_{\text{m}} = - iQ/m_{\text{eff}} \Omega_{\text{m}}^2$ is the mechanical susceptibility at resonance. We introduce a complex fit parameter $\eta$ in order to scale the overall magnitude of the force and to include a phase offset. In Fig. \ref{fig4}c and \ref{fig4}d we plot the normalized RMS amplitude $A_0/Q$ and the phase $\Phi_0$, respectively, using the simulated acceleration presented in Fig. \ref{fig3} and the theoretical curve of the lifetime $\Gamma_{\text{I}}^{-1}$ in Fig. \ref{fig2}b. 
Note that $\eta$ is determined by fitting only to the $[010]$-data, for which we expect the force due to deformation-potential coupling to be dominant. This results in $|\eta| = \SI{0.39+-0.07}{}$ and $\text{arg}(\eta)/\pi = \SI{- 0.55+-0.02}{}$. The same value for $\eta$ is then used for the model curve of the membrane aligned along direction $[\bar110]$, in which case the piezoelectric force prevails. This method is crucial in order to distinguish the strength and sign of the two forces, which constitutes an important aspect of our work.

The response measurements and our model follow the same overall behavior: for $V_\text{b} > \SI{-0.5}{\volt}$ and relatively short lifetimes of the EHPs we get low amplitudes $A_0$ corresponding to a small number of indirect EHPs. For $V_\text{b} \approx \SI{-0.5}{\volt}$, the decay rate $\Gamma_{\text{I}}$ nearly matches the mechanical frequency of $\Omega_{\text{m}}\approx 2\pi \times \SI{1.52}{\mega \hertz}$, resulting in an increased number of carriers and therefore a large $A_0$.  Most importantly, the phase $\Phi_0$ drops abruptly, evidencing a significant delay of the membrane motion with respect to the optical drive. When $V_\text{b} < \SI{-0.5}{\volt}$, the phase decreases further while the amplitude is nearly constant. 
With regard to the membrane orientations investigated here, we see that the maximum displacement of the membrane aligned along $[\bar110]$ is approximately four times larger than the one for the membrane rotated by \SI{45}{\degree}. We also observe another important feature: the difference in phase between the two orientations is approximately $\pi$, revealing an opposite sign of the carrier-mediated force as expected from our simulation (see Fig. \ref{fig3}).

To explain discrepancies between data and theory, 
we first consider the screening of carriers \cite{NSE2000prb} as a possible source of systematic error. We estimate the steady-state number of carriers $\bar{n}_\text{I} = \alpha \bar{P}/ \hbar \omega_\text{L} \Gamma_\text{I} \pi r_\text{s}^2$, where $\bar{P} = P_0 = \SI{0.5}{\micro \watt}$ is the average power of the modulated incident light and $r_\text{s}$ the spot radius shown in Fig. \ref{fig2}d. A maximal density of $\bar{n}_\text{I}  \approx \SI{4e9}{\per \cm \squared}$ is found for $V_\text{b} = \SI{-2}{\volt}$ and $\Gamma_\text{I}^{-1} = \SI{750}{\nano \s}$, resulting in a carrier-induced electric field of $q \bar{n}_\text{I} / \epsilon_0 \epsilon_\text{r} = \SI{56}{\kilo \volt \per \meter}$, where $\epsilon_\text{r} \approx 13$ is the relative permittivity of GaAs and $q$ the elementary charge. This electric field is nearly two orders of magnitude smaller than the one created by the applied bias voltage of $V_\text{b} = \SI{-2}{\volt}$. Therefore, we conclude that carrier-induced screening is negligible. 

We suspect that a more complex description of the EHP states than our simplistic three-level scheme is necessary to understand the disagreement in the above-mentioned voltage interval. This may include higher order EHP states \cite{DTLS2015oe} and dark states \cite{PKTLGPG2010np, srkp1997prb}, as well as non-radiative recombination. Furthermore, we note that surface recombination of carriers at the membrane edges is not included in our model and may significantly influence the carrier dynamics. Outliers at around $V_\text{b} = \SI{1}{\volt}$ are likely due to diode forward current, introducing additional charges in the CQWs.

As indicated in Fig. \ref{fig4}c, the carrier-mediated forces studied here are significantly stronger than radiation pressure. The latter is estimated for the incident optical power of amplitude $P_0 = \SI{0.5}{\micro \watt}$ used in the experiment and results in an RMS displacement of $A_{\text{rad}} = \sqrt{2} r P_0 Q/m_{\text{eff}} \Omega_{\text{m}}^2 c \approx \SI{1.32}{\pico \meter}$ (or $A_{\text{rad}}/Q \approx \SI{0.051}{\femto \meter}$), where $r = 0.66$ is the membrane reflectivity at a wavelength of \SI{880}{\nano\m} and $m_{\text{eff}} = \SI{340}{\pico \gram}$. In comparison, the maximum value measured at $V_\text{b} = \SI{-1.2}{\volt}$ is $A_0 = \SI{1.314+-0.002}{\nano \meter}$ (or $A_0/Q = \SI{61.6+-0.3}{\femto \meter}$), which is three orders of magnitude larger. The membrane's geometry and heterostructure can be optimized for even larger carrier-mediated displacement, for example by reducing the membrane thickness and adjusting the position of the CQWs.
With regard to the cavity optomechanics approach, it should be noted that the carrier-mediated forces presented here rely on absorption of light. This prohibits the use of high-finesse optical cavities, which typically enhance the forces in radiation pressure based optomechanical systems and thereby enable large dispersive coupling.

\section{Conclusion}

We have examined the effects of indirect EHPs in CQWs on the mechanical motion of a free-free nanomembrane via the deformation-potential and piezoelectric forces. Despite discrepancies between our model and data, our theoretical description reproduce important features of the optically driven response measurements, revealing in particular that both strength and delay of the carrier-mediated forces are tuned via the lifetime of the EHPs, which in turn is controlled by a bias voltage across the CQWs. 
We hope that our work will stimulate research into novel regimes and interactions in quantum optomechanics involving, for example, polaritons \cite{KLS2014prl, RBFJL2014prb, Restrepo2015sr, Zhou2016pra} or active cavities of semiconductor lasers \cite{Czerniuk2014natcom, Yang2015sr}. In combination with an optical cavity (see Appendix \ref{app:integratedcav}), the device presented in this work may implement dynamical backaction based on the deformation-potential in semiconductors.


\section*{Acknowledgements}

We gratefully acknowledge Ignacio Wilson-Rae, Dalziel Joseph Wilson, Andreas N\ae sby Rasmussen, Koji Usami and Rapha\"el S. Daveau for valuable assistance at the early stage of this work and Javier Miguel-S\'{a}nchez for growing the wafer. This work was supported by the DARPA project QUASAR, a Sapere Aude starting grant and postdoctoral grant (grant no. 4184-00203) from the Danish Council for Independent Research, Q-CEOM (grant no. 638765), the Villum Foundation, the Carlsberg Foundation, as well as the European Union Seventh Framework Program including the projects SIQS, iQUOEMS, and the ERC grants INTERFACE. This project has received funding from the European Research Council (ERC) under the European Union’s Horizon 2020 research and innovation programme (grant agreement No 787520).


\section*{Author contributions}

A.B. and L.M. contributed equally to this work. A.I., E.S.P. and S.S. proposed the experiment. E.S.P., S.S., A.S. and P.L. supervised the project. S.S. and P.T. designed the device and modeled the carrier lifetime. A.B. and L.M. conceived the theoretical model with input from A.S., while L.M. conducted the finite element simulations. L.M. and T.P. fabricated the sample. G.K. and A.B. performed the photoluminescence measurements. A.B. carried out the interferometric measurements and analyzed the data. A.B. and L.M. wrote the manuscript with input from all authors.


\appendix

\section{Device fabrication}

The membranes are grown by molecular beam epitaxy on a $[001]$ GaAs wafer, separated from the substrate by a \SI{2}{\micro \meter} thick Al$_{0.75}$Ga$_{0.25}$As sacrificial layer.
The electrical gates to the p-i-n diode are fabricated first. An access via to the n-doped GaAs layer is defined by ultra-violet lithography and subsequently etched in a reactive ion etching (RIE) plasma (BCl$_3$/Ar 1:2). This step also defines a mesa with an area of $2 \times \SI{1}{\milli \meter^2}$ that isolates the p-layer from the rest of the chip. A precise ($\pm \SI{10}{\nano \meter}$) end-point detection based on laser interferometry allows stopping the etch process a few tens of nanometers above the n-layer to achieve optimal and reproducible ohmic contacts. Square pads ($180 \times \SI{180}{\micro \meter^2}$) are patterned in a negative photo-resist on the exposed n-layer. The metal sequence Ni/Ge/Au/Ni/Au (5/40/60/27/\SI{100}{\nano \meter}) is deposited by electron gun evaporation and lifted-off in acetone. The contacts are further annealed in a rapid thermal annealer at \SI{420}{\celsius}. A similar procedure is performed to deposit p-type contacts on the surface of the membrane. Here, a Au/Zn/Au/Ti sequence (20/50/150/\SI{7}{\nano \meter}) is deposited by thermal evaporation. The last Ti layer is evaporated to increase the photo-resist adhesion in subsequent steps. The contacts are also annealed for 1 minute at \SI{420}{\celsius}. To protect the contacts from photo-corrosion (due to the presence of doped layers) and galvanic erosion at the metal-semiconductor interfaces, several protection rings made of a \SI{500}{\nano \meter}-thick photoresist (AZ1505) are fabricated around the pads and hard-baked to make them adhere to the chip permanently.  

\begin{figure}[t!]
\includegraphics[width=0.75\linewidth]{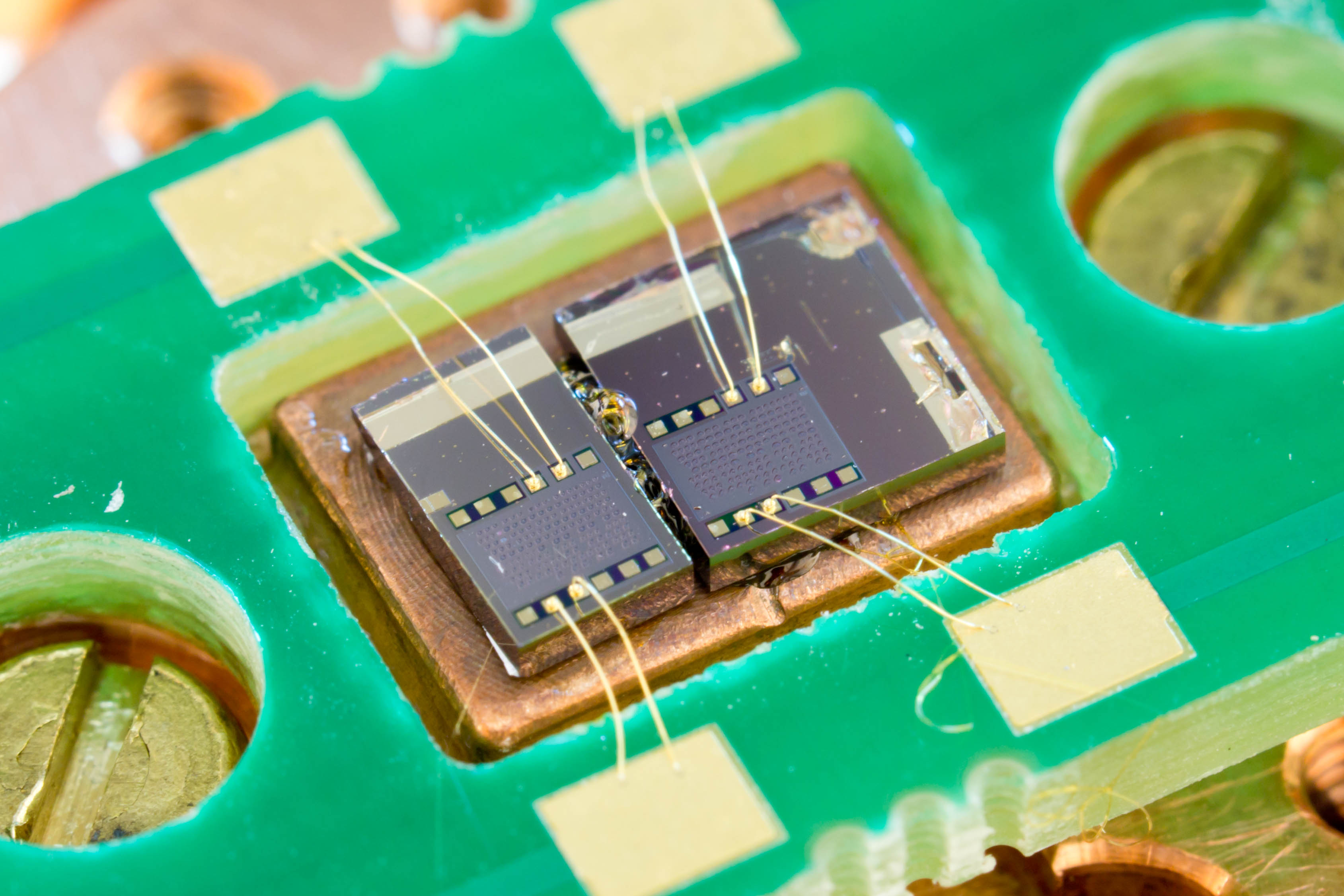}
\caption{\textbf{Fabricated device.} Photograph of two samples mounted on a copper pad with electrical contacts to a PCB. Each sample contains 140 free-free nanomembranes.}
\label{fig:devicefab}
\end{figure}

The membranes are fabricated with a soft-mask method whose details have been given in a previous work \cite{Midolo2015n}. A \SI{550}{\nano \meter}-thick electron-beam resist (ZEP520A) is spin-coated on the entire sample and various free-free membrane designs are patterned by electron-beam lithography aligned to the mesa and parallel or 45 degrees rotated from the GaAs cleavage planes. The membranes are finally etched by inductively coupled plasma RIE (BCl$_3$/Cl$_2$/Ar 3:4:10) and undercut by hydrofluoric acid. A final cleaning step in hydrogen peroxide is performed to remove photoresist residues. This step etches away a small amount of AlGaAs in the membrane, delineating the layout structure as shown in Fig. \ref{fig1}b. The samples are processed in a CO$_2$ critical point drier to avoid collapsing of the membranes on the substrate due to capillary forces.
After processing, the sample is mounted on a copper pad and wire-bonded to a printed circuit board (see Fig. \ref{fig:devicefab}).

\section{Photoluminescence measurements}
\label{app:plmeasurements}

Photoluminescence (PL) measurements are performed with a sample placed inside a cryostat at a temperature of $T = \SI{12}{\kelvin}$. A microscope objective focuses laser light from a picosecond-pulsed diode laser at a wavelength of \SI{785}{nm} to a spot radius of $\sim \SI{0.75}{\micro \meter}$ at the center of a membrane. The PL from the CQWs is collected by the same objective and transmits through a longpass dichroic mirror with a cutoff wavelength of \SI{875}{nm}. Afterwards, the light is focused onto a CMOS camera, creating an image of the membrane which we use to determine the carrier density distribution. 

\begin{figure}[t]
\includegraphics[width=0.6\linewidth]{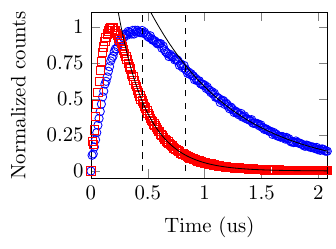}
\caption{\textbf{Time-resolved PL measurements.} Normalized photon counts at the indirect transition wavelength as a function of time after picosecond-pulsed excitation at \SI{785}{\nano \meter}. At bias voltages of $\SI{-2}{\volt}$ (blue circles) and \SI{-1}{\volt} (red squares) we extract lifetimes $\Gamma_\text{I}^{-1}$ of \SI{749.5+-0.4}{\nano \s} and \SI{256.6+-0.1}{\nano \s}, respectively, where the statistical errors are taken from the individual fits. The black solid lines are exponential fits while the dotted lines indicate the start times of the fits.}
\label{fig:lifetime}
\end{figure}

Alternatively, the light can be sent to a spectrometer with a resolution of \SI{50}{\pico \meter}, where we filter out photons from radiative recombination via the indirect transition. An avalanche photodiode is used to perform time-resolved measurements of the filtered photons. As exemplified in Fig. \ref{fig:lifetime}, the photon counts quickly build up after the pulse as generated carriers decay into the indirect state $\left|\text{I}\right\rangle$ via acoustic phonons. Then, the counts slowly decrease. To extract the lifetime of the indirect transition $\Gamma_\text{I}^{-1}$ for different bias voltages $V_\text{b}$, we fit exponential functions to the tail of the curves and thereby minimize contributions from the buildup.

\section{Simulations}
\label{app:simulations}

To extract the magnitude and sign of the relevant carrier-induced forces involved in the experiment, a finite element simulation of the mechanical properties of free-free membranes has been carried out. The geometry used for the simulations comprises one quarter of the membrane and a \SI{5}{\micro \meter} undercut ring. The structure is fixed on the outer ring and symmetric boundary conditions are used to identify the modes of interest. Additionally, the membrane mesh is deformed according to the profile measured by white light confocal microscopy (see Fig. \ref{fig:toporefl}) to take into account the effect of buckling.
Both GaAs and AlGaAs material properties (elastic and piezoelectric matrices) are anisotropic according to the Zincblende structure. These quantities are summarized in Table \ref{tab:structural}.
\begin{table}[b!]
  \centering
    \begin{tabular}{ llrl}
			\toprule
			Material & Component & Value\\
			\hline
			\noalign{\vskip 1mm} 
			GaAs & $D_{11}=D_{22}=D_{33}$ & \SI{121.56}{\giga \pascal} \\
			& $D_{12}=D_{13}=D_{23}$& \SI{54.54}{\giga \pascal}\\
			& $D_{44}=D_{55}=D_{66}$& \SI{6.5}{\giga \pascal}\\
			& $e_{14}=e_{15}=e_{16}$& \SI{-0.16}{\coulomb \meter^{-2}}\\
			& $\rho$ & \SI{5307}{\kilogram \per \meter \cubed}\\
			\noalign{\vskip 1mm} 
			\hline
			\noalign{\vskip 1mm} 
			Al$_{0.4}$Ga$_{0.6}$As& $D_{11}=D_{22}=D_{33}$ & \SI{119.36}{\giga \pascal}\\
			& $D_{12}=D_{13}=D_{23}$& \SI{55.08}{\giga \pascal}\\
			& $D_{44}=D_{55}=D_{66}$& \SI{5.9}{\giga \pascal}\\
			& $e_{14}=e_{15}=e_{16}$& \SI{-0.186}{\coulomb \meter^{-2}}\\
			& $\rho$ & \SI{4696}{\kilogram \per \meter \cubed}\\ \noalign{\vskip 1mm} \hline
    \end{tabular}
\caption{\textbf{Structural parameters used in the simulation at a temperature of $T = \SI{10}{\kelvin}$.} The indices from 1 to 3 refer to the crystal orientation whereas from 4 to 6 to shear effects around the corresponding crystal orientations. $D_{ij}$ is the elastic coefficient, $e_{ij}$ the stress-charge piezoelectric coefficient, and $\rho$ the density.}
	\label{tab:structural}
\end{table}

The deformation-potential force is included as an internal hydrostatic stress $\sigma^\text{dp}$ acting on the layers of the CQWs only:
\begin{equation}
\sigma^\text{dp}_{ij} = \delta_{ij} \tilde{n}_\text{I}(\Omega_\text{m}) \left(E^\text{dp}_\text{e}\Psi_\text{e}(z)-E^\text{dp}_\text{h}\Psi_\text{h}(z) \right),
\end{equation}
where $i=1,2,3$ are the principal directions of GaAs and $E^\text{dp}_\text{e} = \SI{-5.3}{\electronvolt}$ and $E^\text{dp}_\text{h} = \SI{+2.7}{\electronvolt}$ are the deformation-potential energies in the conduction and valence band, respectively \cite{matsuda2005acoustic}. $\tilde{n}_\text{I}(\Omega_\text{m})$ is the complex carrier density distribution of indirect electron-hole pairs (EHPs) modulated at the mechanical frequency $\Omega_\text{m}/2\pi \approx \SI{1.6}{\mega\Hz}$ and calculated numerically using Eq. (\ref{eq:diffusiondensity}). Here, we assume a carrier generation rate
\begin{equation}
	G = \frac{\alpha}{\hbar \omega_\text{L}} e^{-\frac{x^2 + y^2}{2 r_\text{s}^2}} I_0 (\sin{(\Omega_\text{m}t)}+1),
\end{equation}
where the amplitude of the modulated power $P_0 = \pi I_0 r_\text{s}^2/4 = \SI{0.5}{\micro\watt}$ and the beam spot radius $r_\text{s} = \SI{1}{\micro\m}$.

The electron and hole profiles $\Psi_\text{e,h}(z)$ along the $z$-axis are approximated by two gaussians peaked on the location of the CQWs, reflecting the combined electron-hole wavefunction of the indirect carriers:
\begin{equation}
\Psi_\text{e,h}(z)=\frac{1}{2\sqrt{2\pi}r_\text{w}}e^{-\frac{(z-z_\text{e,h})^2}{2 r_\text{w}^2}},
\end{equation}
where $z_\text{e,h}$ are the positions of the quantum wells, and $r_\text{w}= \SI{2}{\nano \meter}$ is the well confinement radius.  

Piezoelectric forces are modelled assuming the modulated distribution of surface charges:
\begin{equation}
\sigma = \pm \tilde{n}_\text{I}(\Omega_\text{m}) q.
\end{equation}
The sign depends on whether holes or electrons are considered and $q$ is the elementary charge. More specifically holes are located at the position of the lower quantum well, whereas electrons are on the upper quantum well as expected from the eigenvalue solution of the Schr\"{o}dinger equation.

Our model allows us to predict the mechanical response of each mode (and its sign) as a function of the type of force, the membrane orientation, and the spot radius.
The modes are calculated solving an eigenvalue problem. A modal reduction technique is used to calculate mass, stiffness matrices and force vectors of the free-free mode. The reduced ordinary differential equation for the displacement $w$ of a mode is given by
\begin{equation}
\ddot{w}+\frac{\Omega_\text{m}}{2Q}\dot{w}+\Omega^2w=\frac{F_\text{eh}}{m_\text{eff}}.
\label{eq:eqmt}
\end{equation}
Here, $\Omega_\text{m}$ is the mechanical frequency of the mode, $Q$ the quality factor, $F_\text{eh}$ the input vector (or effective force), and $m_\text{eff}$ the mode effective mass.
To calculate the effective masses we normalize the eigenmode's total displacement $\phi(r)$ such that its range is between -1 and 1 (dimensionless). This also ensures that the displacement $w$ has the unit of meter.
The effective mass is given by
\begin{equation}
m_\text{eff}=\int dV \rho (r) \left|\phi (r) \right|^2,
\end{equation}
where the integral is taken over the entire computational domain. 
For the membrane discussed in this article and the free-free eigenmode we find $m_\text{eff} = \SI{340}{\pico \gram}$.

The quality factor $Q$ in Eq. (\ref{eq:eqmt}) is introduced phenomenologically as a loss factor in the material ($1/Q^*=\SI{2.4e-5}{}$) according to Cole et al. \cite{Cole2011nc}. The clamping losses have been neglected since the free-free design ensures a very limited phonon tunneling into the bulk. From the experiments (see Fig. \ref{fig1}c) we conclude that, when thermoelastic damping is absent, $Q$ is in fact on the order of $Q^*$.

The figure of merit for the force is given by the overlap integral between the input stress and the free-free mode. This quantity is summarized by the complex effective acceleration impressed on the mechanical mode $a_\text{eh}(\Omega_\text{m}) = F_\text{eh}(\Omega_\text{m})/m_\text{eff}$, where the real and complex parts reveal the in-phase and out-of-phase drive, respectively. 
We can compare these values for the case of a piezoelectric force and deformation-potential for different orientations. This is done in the model by rotating the reference frame of the geometry compared to the material. The results are shown in Fig. \ref{fig3}.

\section{Identification of membrane orientation}

\begin{figure}[h!]
\includegraphics[width=\linewidth]{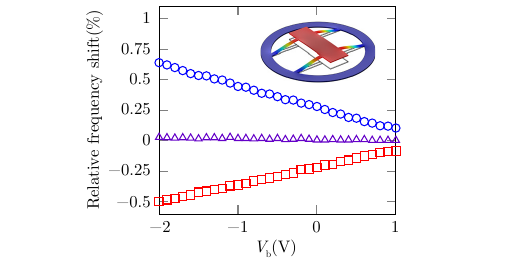}
\caption{\textbf{Identification of membrane orientation.} Relative frequency shift of drum mode due to piezoelectric effect as a function of bias voltage $V_\text{b}$ for three different membranes. Due to the slope of the linear shift, we identify the membrane orientations along $[\bar110]$ (blue circles), $[010]$ (purple triangles), and $[110]$ (red squares). A simulation of the mode shape is shown as an inset.}
\label{fig:orientation}
\end{figure}

\begin{figure}[b]
\includegraphics[width=\linewidth]{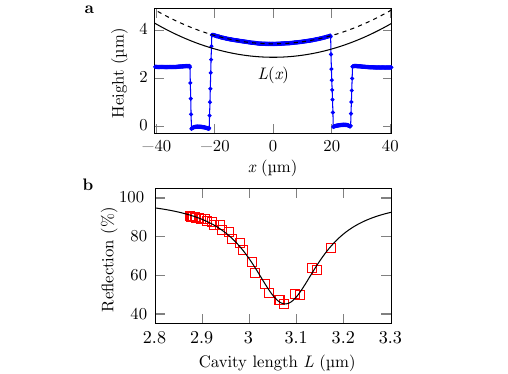}
\caption{\textbf{Characterization of integrated cavity.} (\textbf{a}) Measured topography (blue dots) along the $x$-axis showing buckling and symmetric membrane deformation. A parabolic fit (dashed line), from which the membrane thickness of $\SI{562}{\nano \meter}$ is subtracted, yields the cavity length $L(x)$ (solid black line). (\textbf{b}) Reflection from the membrane as a function of $L$ at a wavelength of 880 nm as measured (red squares) and fitted with a Lorentzian (solid black line). From the fit we estimate a finesse of $\mathcal{F} = \SI{2.48+-0.09}{}$.}
\label{fig:toporefl}
\end{figure}

The inset of Fig. \ref{fig:orientation} shows a simulation of a drum mode, where the entire free-free membrane displaces vertically. The frequency of this mode is $\sim \SI{900}{\kilo \hertz}$ and depends on the length of the four thin beams, by which the central plate is suspended. When a bias voltage $V_\text{b}$ is applied, shear stress described by the piezoelectric stress tensor causes a length change of the beams, resulting in a frequency shift. We simulate the effect and find that a membrane oriented along the crystallographic axis $[\bar{1}10]$ experiences a linear shift with negative slope as a function of $V_\text{b}$, while for a membrane oriented along the orthogonal direction $[110]$ the slope is positive. When aligned with the direction $[010]$ or $[100]$, a much smaller frequency shift occurs. As shown for three different cases in Fig. \ref{fig:orientation}, the behavior is observed experimentally and used to determine the membrane orientation.
Notice that for all membranes we choose the frequency shift to be zero at $V_\text{b} = \SI{1.52}{\volt}$, where the bias voltage cancels out the built-in field and no piezoelectric effect is expected.

\section{Integrated optical cavity}
\label{app:integratedcav}

Our membranes are suspended above an integrated planar distributed Bragg reflector (DBR), which is optimized for the wavelength range of the indirect EHP transition (\SI{900}{\nano \meter} to \SI{950}{\nano \meter}). At $\lambda = \SI{880}{\nano\meter}$ the DBR's reflectivity is estimated via the transfer matrix method to be \SI{76}{\%}. DBR and membrane constitute a low finesse optical microcavity, which we characterize by topography and reflection measurements. Figure \ref{fig:toporefl}a shows the height profile measured with a white light confocal microscope across the long side of a membrane ($x$-axis). As can be seen from this example, our membranes buckle and deform symmetrically in accordance with stress release of the heterostructure after underetching. From a parabolic fit to the deformed membrane measured here we determine the position-dependent cavity length $L(x)$ and, by calculating its second derivative at $x = 0$, the radius of curvature in the center to be \SI{577.8+-1.9}{\micro \meter}. 

Next, we map the reflection along a centered line on the membrane using a focused laser beam at a wavelength of $\SI{880}{\nano\meter}$. For each data point, the position $x$ is translated into a cavity length via $L(x)$. The result, shown in Fig. \ref{fig:toporefl}b, features an optical resonance at $ L \approx \SI{3.1}{\micro \meter}$.
From a Lorentzian fit to the data we extract a full width half maximum of $\SI{177.0+-6.5}{\nano \meter}$. Assuming a distance between two resonances $\lambda/2 = \SI{440}{\nano \meter}$, we further estimate an optical finesse of $\mathcal{F} = \SI{2.48+-0.09}{}$.

The integrated cavity could be used to realize moderate optomechanical cooling and amplification of the membrane motion via carrier-induced forces. For instance, red and blue detuning from the cavity resonance could be controlled via the cavity length, while maintaining the above-bandgap laser light at 880 nm as used in our work. Indeed, we have observed amplification and self-oscillations showing a dependence on the bias voltage $V_\text{b}$. A detailed study of these observations is needed.


\bibliography{cqw}

\end{document}